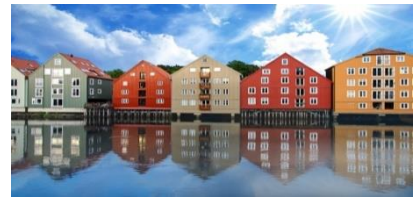

## Paper Information

| | |
|---|---|
| Paper number | 1120 |
| Paper title | Optimal and Coordinated Voltage Control: Case Study on a 132 kV Norwegian Grid Subsystem |
| Study Committee | SC C2 – Power system operation and control |
| Preferential subject | PS2: Technologies supporting the power grid for energy transition to carbon neutral energy production |
| Authors | Hugo Rodrigues DE BRITO, Daniel Simon BALTENSPERGER, Kjetil Obstfelder UHLEN |
| Affiliations (optional) | Norwegian University of Science and Technology (NTNU) |
| Country | Norway |
| Email address | hugo.r.de.brito@ntnu.no |

## Summary

Grid reactive power management through voltage regulation is one of the great challenges faced by transmission system operators (TSOs) worldwide in their efforts to facilitate energy transition and electrification. In Norway, several new services and coordinated actions are being developed to maintain power balance and congestion management. However, without proper control of voltage levels and reactive power flows, utilization of the available grid capacity is bound to be suboptimal regarding power losses.

In this context, the Norwegian TSO has expressed great interest in novel solutions for power loss minimization via optimization of reactive power resources using existing grid assets. This translates into both preventive measures, potentially postponing the need for network reinforcements, and corrective measures for voltage instability-prone system areas. As voltage regulation is carried out remotely through regional control centres, it is paramount for the proposed strategies to be hierarchical in nature. This type of control is typically divided into three layers: primary (PVR), secondary (SVR) and tertiary (TVR) voltage regulation. While PVR is mainly automatic, most of the coordination amongst higher levels is still performed manually by operators today. It is foreseen that with larger and less predictable power flow variations, coordinated and automatic regulation is expected to become the norm for all layers.

This work presents a framework for dynamic performance assessment of the higher layers in the hierarchical scheme, with case studies applied to specific areas of the Norwegian grid. Unlike the PVR level, the SVR and TVR levels are not tuned to a single device at a time, handling instead several reactive power resources available within a control zone including generator units, static VAr compensators and others. Proper SVR-TVR coordination for realistic transmission systems is a challenging topic at the core of many ongoing discussions in voltage control literature.

The SVR design is implemented using a parallel approach, based on conventional proportional-integral controllers for simultaneous pilot bus setpoint tracking and reactive power sharing adjustments. This structure aims to replicate the standard configuration of real-life SVR schemes from certain countries. The TVR design, on the other hand, is implemented as a centralized optimizer through the formulation of an optimal power flow (OPF) problem, with the objective of determining reactive power setpoints towards active power loss reduction. For this purpose, the OPF tackles TSO-owned shunt-controlled elements and assets belonging to generator companies. In both designs, special focus is placed on practical considerations from the system operator perspective, since this research is also aimed at simplifying daily control centre routines.

Dynamic simulation results concern a 21-bus equivalent of a 132 kV network model that accurately represents a Norwegian grid subsystem. Case studies address daily grid operation with real-life load demand and wind power generation profiles, showing that the proposed strategy is effective not only to minimize total active power losses as much as possible within system-wide limitations, but also to maintain adequate voltage profiles and reactive power flows. Findings pertaining to this work showcase the benefits of applying hierarchical voltage regulation layers as an asset to day-to-day control center management of a realistic transmission network.

# 1    Introduction

Worldwide electrification ambitions are bound to reach new heights with each passing year, particularly as the largest economies recover from the disruption of 2020. Recent findings from the International Energy Agency (IEA) indicate that year-on-year global electricity demand increases by 5 % – over 1200 TWh – on average. At the same time, renewables-based generation tends to outpace such growth, already accounting for over 50 % of the supply increase required to meet the updated demand in 2023 [1].

The Norwegian electricity system presents a similar trend: annual power production is only expected to intensify throughout the upcoming decade, spearheaded by a dependable hydropower sector currently representative of 88 % of the country's total supply [2]. On the other hand, rising line power flows bring about reduced net export margins and lower transfer capacity within the Nordic power grid. Such limitations are further aggravated by active power losses at the transmission level, which account for a progressively larger share of the total network energy costs [3].

In this context, Statnett – the transmission system operator (TSO) in Norway – has expressed great interest in novel solutions for loss minimization via optimized reactive power management of existing grid assets. As it is known, reactive power flow and voltage magnitude are strongly coupled at the transmission level, meaning that strategies based on voltage control techniques are especially desirable [4]. Studies catering to such TSO needs are still on-going and widespread, ranging from the Southern [5] to the Northern [6] extremes of the country.

Although voltage regulation is traditionally carried out manually within regional control centers, a major shift to coordinated and hierarchical structures has been observed over the last decades. Coordinated voltage control is usually divided into three hierarchical levels: primary voltage regulation (PVR), secondary voltage regulation (SVR) and tertiary voltage regulation (TVR). Each level is decoupled from others in terms of action zones and time scales so as to avoid undesired interactions among device controllers [7].

As the intermediate control level, the SVR layer is responsible for maintaining an adequate voltage profile at buses within a predefined control area, which might include several generator units, flexible ac transmission systems (FACTS) and other reactive power resources. The upper TVR layer, on the other hand, acts upon the entire power grid and is typically associated to optimal power flow (OPF) solutions obtained at discrete time intervals. SVR and TVR schemes are a staple of literature on voltage stability, with several successful implementations in both simulated and real-life settings.

Reference [8] summarizes the major existing SVR-TVR configurations. The authors of [9] and [10] opt to combine the SVR layer with voltage stability indices and TVR, respectively, as a means to account for network vulnerabilities in an optimized manner. Practical experiences of hierarchical voltage control are emphasized in [11], where its merits as an ancillary service to the Spanish TSO are highlighted, and in [12], where its fully automated nature is proof-tested in a real HV network managed by the Croatian TSO.

These real-life examples not only facilitate the understanding of hierarchical operation philosophies, but also help identify possible improvements applicable to specific areas of the Nordic power grid. As pointed out in [8], a small SVR prototype has been implemented within the TSO's regional control center for Southern Norway in the 2000s. Around 20 years later, no innovation nor concrete expansion upon this idea has been put into practice in the country.

The outline and contributions of this paper are summarized as follows: Section 2 presents key design considerations for the SVR control layer employed in case studies, focusing on modelling and objectives of a PI-based parallel scheme; Section 3 describes the OPF-based TVR control layer by detailing its problem formulation as an expansion of the classical interior-point method; Section 4 presents the 132 kV Norwegian grid subsystem which forms the basis of assessment of the combined SVR-TVR framework; Section 5 summarizes the main dynamic simulation results for bus voltage profiles, reactive power flows and active power losses as a means to showcase the merits of the proposed approach; Section 6 concludes the paper.

## 2   The Secondary Voltage Regulation Layer

To properly define SVR capabilities, it is important to have in mind the fundamental distinctions among coordinated control layers. Fig. 1 shows the spatial-temporal decoupling that characterizes the primary, secondary and tertiary hierarchical voltage regulation levels.

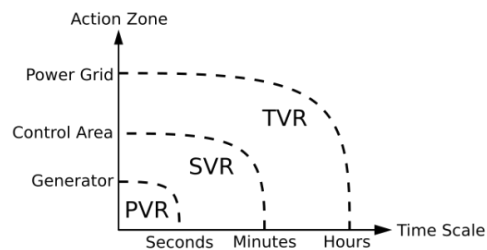

*Figure 1 – Spatial-temporal decoupling in a three-level hierarchical voltage control scheme*

As illustrated, each layer is associated to a different action zone and time scale: PVR action is typically at a generator level, concerning automatic voltage regulators (AVRs), and tends to be the fastest with a range of fractions of a second to some seconds; SVR action involves a predefined control area, including several generator units, and presents slower dynamics with a range of some seconds to some minutes; TVR action influences the entire power grid, adjusting the overall system profile through optimization techniques, and is the slowest with a range of some minutes to several hours.

The hierarchical structure mitigates conflicting control objectives which could otherwise cause long-term issues, such as voltage runaway and wear-and-tear of controllers. Communication between layers is bidirectional, carried out by control signals in such a way that the broadest layer always takes precedence over the narrowest one. In this context, the SVR scheme is responsible for maintaining voltage levels at an acceptable operating range within its action zone, a typically small control area predefined to be electrically distant from other SVR control areas. That way, impact of local corrective actions on external voltage behavior is minimized.

Regardless of design philosophy, the conventional SVR structure can be divided into two main components: the central pilot bus controller, a dispatcher of corrective signals for pilot bus setpoint tracking; and a set of distributed power plant controllers, coupled to each participating reactive power source for individual adjustments. A simple realization of both of these controllers requires proportional-integral (PI) blocks as well as feedback signals from the pilot bus and from the PVR layer within the hierarchical scheme.

In the parallel SVR approach, the controller blocks are not connected to each other in any way, meaning that they work independently towards fulfilling specific control objectives. Thus, their outputs are separately fed back to the PVR layer, where the resulting signal triggers the necessary corrective measures. These are smoothly carried out over time in accordance with the medium-term SVR dynamics. Fig. 2 shows the parallel SVR scheme with PI-based components, focusing on generators as designated reactive power sources (i.e., AVRs make up the PVR layer). Details on each controller's functionalities and parameters are as follows:

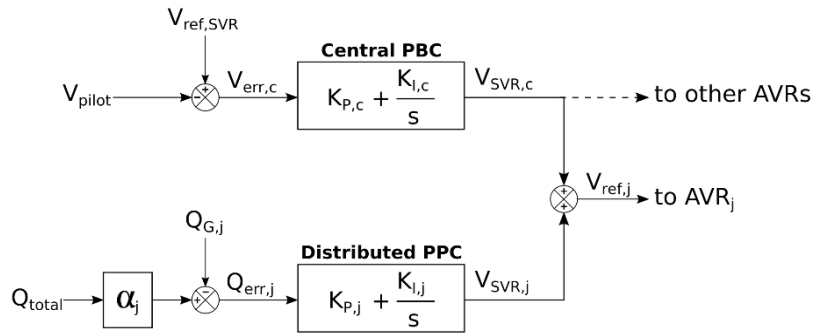

*Figure 2 – Parallel SVR scheme divided into PI controllers*

- Central pilot bus controller (PBC): monitors and corrects $V_{pilot}$ according to a predefined $V_{ref,SVR}$, generating the error signal $V_{err,c}$. Necessary adjustments are shared through a $V_{SVR,c}$ signal sent to all participating AVRs, which is the output of the PI block with $K_{P,c}$ and $K_{I,c}$ gains;
- Distributed power plant controller (PPC): provides individual reactive power adjustments for generators, based on a predefined $\alpha_j$ usually proportional to each machine's power rating. This factor scales $Q_{total}$ for comparison with $Q_{G,j}$, generating the error signal $Q_{err,j}$. The net result of this action is a $V_{SVR,j}$ signal sent to the respective $AVR_j$, which is the output of the PI block with $K_{P,j}$ and $K_{I,j}$ gains.

This configuration can be fully described by the PI control law for $V_{ref,j}$, the reference voltage which serves as input to the AVR for generator terminal voltage adjustments. Such law can be mathematically derived in the frequency domain as ($s$ being the Laplace operator):

$$V_{ref,j}(s) = \left(K_{P,c} + \frac{K_{I,c}}{s}\right) V_{err,c}(s) + \left(K_{P,j} + \frac{K_{I,j}}{s}\right) Q_{err,j}(s) \qquad (1)$$

According to (1), the closed-loop relationship established by the parallel SVR is essentially a sum of two first-order transfer functions, which is informative of its dynamic performance. Furthermore, it is possible to express the control objectives in the time domain as ($t$ being the time parameter):

$$\lim_{t\to\infty} V_{pilot}(t) = V_{ref,SVR}, \qquad \lim_{t\to\infty} \left[\frac{Q_{G,j}(t)}{Q_{total}(t)}\right] = \alpha_j \qquad (2)$$

From (2), it is worth noting that the SVR aims both to track the pilot bus setpoint and to ensure a fair reactive power sharing among participating generators with its central and distributed controllers. In the parallel approach, these objectives are complementary not only due to independent controller operation, but also because voltage magnitudes and reactive power flow are strongly coupled at the transmission level.

Despite its simplicity, the parallel structure of Fig. 2 is nonetheless conceptually similar to certain real-life SVR schemes. It constitutes the standard configuration for SVR studies in countries such as Brazil [13], Malaysia [14] and Colombia [15]. The main difference is the assumption of continuous operation, whereas practical SVR implementations are discrete in nature with sampling rates in the range of seconds.

## 3  The Tertiary Voltage Regulation Layer

In this paper, the TVR layer has the main goal of minimizing active power losses through a traditional deterministic OPF formulation, namely the interior-point method. The proposed framework is summarized in the standard writing of optimization problems as follows:

$$\min_{Q_G, V} \sum_{(i,j) \in \mathcal{M}} (P_{ij} + P_{ji}) \tag{3}$$

$$s.t. \ g(u,x) \leq 0, \quad h(u,x) = 0$$

Where $Q_G$ is the vector of generator reactive power outputs, $V$ is the vector of nodal voltage magnitudes, $\mathcal{M}$ is the set of all nodes and $P_{ij}, P_{ji}$ are the line active power flows. The vector $x$ contains all states, such as load bus parameters, and the vector $u$ contains all control variables, such as generator terminal parameters. Thus, the constraint $g(u,x)$ is meant to represent a set of inequalities (line loading limits, voltage magnitude thresholds, etc.) and the equality constraint $h(u,x)$ comprises the classical set of power flow equations.

Equations (4)-(9) show the constraints used for the tuning of synchronous generators within the system, where $\mathcal{G}$ is the set of all machines. The most important parameters are the user-defined governor droop control $K_{p,k}$ (related to the frequency variation $\Delta f$), the control variable representing small changes in reactive power output $\Delta Q_{G,k}$ and the maximum leading power factor $\varphi_{lead}^{max} = 0.86$ as per the Norwegian grid code. Other parameters stand for initial, current and minimum/maximum active, reactive and complex generator power output, as well as internal machine voltage and synchronous reactance.

$$P_{G,k} = P_{G,k}^0 + K_{p,k}\Delta f, \quad \forall k \in \mathcal{G} \tag{4}$$

$$P_{G,k}^{min} \leq P_{G,k} \leq P_{G,k}^{max}, \quad \forall k \in \mathcal{G} \tag{5}$$

$$Q_{G,k}^{min} \leq Q_{G,k}^0 + \Delta Q_{G,k} \leq Q_{G,k}^{max}, \quad \forall k \in \mathcal{G} \tag{6}$$

$$0 \leq \sqrt{(P_{G,k})^2 + (Q_{G,k}^0 + \Delta Q_{G,k})^2} \leq S_{G,k}^{max}, \quad \forall k \in \mathcal{G} \tag{7}$$

$$-P_{G,k} \tan(\varphi_{lead}^{max}) \leq Q_{G,k}^0 + \Delta Q_{G,k}, \qquad \forall k \in \mathcal{G} \tag{8}$$

$$0 \leq (P_{G,k})^2 + \left(Q_{G,k}^0 + \Delta Q_{G,k} + \frac{V^2}{x_d}\right)^2 \leq \left(\frac{VE_{q,max}}{x_d}\right)^2, \qquad \forall k \in \mathcal{G} \tag{9}$$

Finally, the optimization modelling of all shunt-controlled elements within the system is carried out through (10). This equation accounts for the operation of both static VAr compensators (SVCs) and static synchronous compensators (STATCOMs) in the set of all shunt devices $\mathcal{C}$, whereby it is defined a bounded search space for the optimizer to select reactive power injection or consumption profiles towards the fulfilment of the overall objective. Both the maximum and minimum reactive power flow thresholds can be tuned to properly represent the availability of such devices at a given operational state of the system.

$$\frac{Q_{SVC}^{min}}{(V_i^{min})^2} V_i^2 \leq Q_{SVC,i} \leq \frac{Q_{SVC}^{max}}{(V_i^{max})^2} V_i^2, \qquad \forall i \in \mathcal{C} \tag{10}$$

## 4 The 132 kV Norwegian Grid Subsystem

Fig. 3 illustrates the single-line diagram of the 132 kV Norwegian grid subsystem, which represents a simplified equivalent of a voltage instability-prone region within Norway. This network is comprised of 21 busbars (B1, …, B21), 24 power lines, 13 load centers, two transformers, four hydropower plant generators (G1, …, G4), four wind parks (W1, …, W4) and five shunt-controlled devices – SVCs and STATCOMs.

All generators and shunt-controlled devices participate to some extent in the SVR-TVR hierarchical control scheme, and a two-area division is proposed in order to better replicate the characteristic behavior of the real system. The line interconnection between busbars B10 and B11 is well regarded as a natural choice for area division with respect to grid topology. For the SVR control, busbars B5 and B14 are defined as pilot buses of Area 1 and Area 2, respectively.

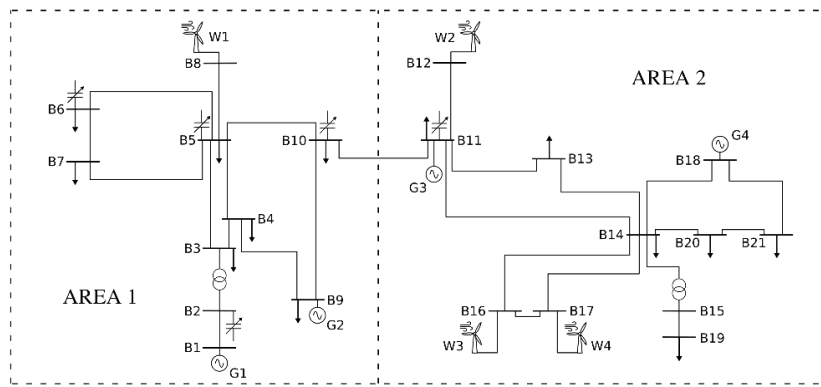

*Figure 3 – Single-line diagram of a 132 kV Norwegian grid subsystem, divided into two SVR control areas*

Load centers are modelled as constant-power type sinks, whereas wind parks operate at unity power factor (i.e., they do not work as reactive power sources). Fig. 4(a) shows the daily load power profiles based on average behavior, which exemplifies a typical power consumption curve for this system peaking in the afternoon period. Fig. 4(b) shows the daily wind active power generation profiles for each wind park, where it can be noticed that, although their capabilities are different, they share a similar overall behavior throughout the day. Load and wind profiles form the basis of the results presented in the next section, concerning bus voltage magnitudes, generator MVAr contributions and total losses throughout the entire network.

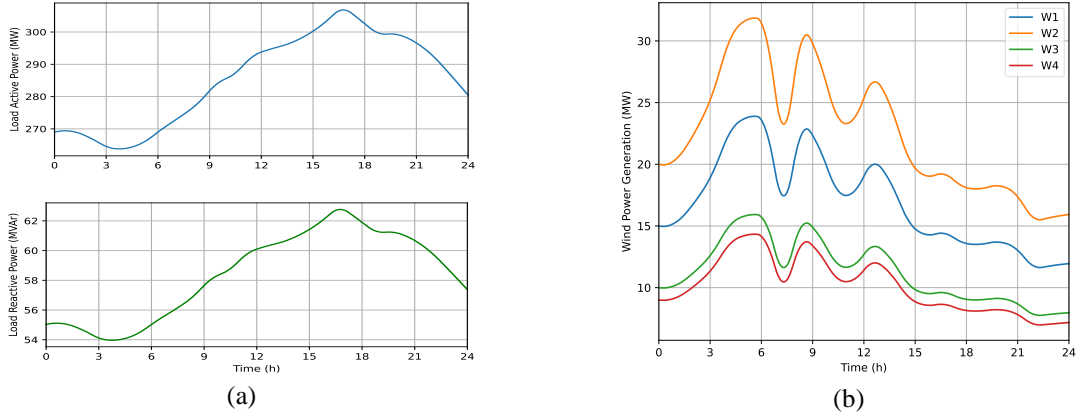

*Figure 4 – Daily profiles of the 132 kV Norwegian grid subsystem for: (a) load active and reactive power demand; (b) wind active power generation.*

## 5   Dynamic Simulation Results

All the plots presented in this section refer to a 24-hour-long simulation of the system in Fig. 3 subjected to the typical behavior pattern of load centers and wind parks of Fig. 4. It is assumed that no contigency events occur, as focus is placed on assessing the impact of the proposed SVR-TVR approach in comparison to a scenario of standard system operation without hierarchical control. The optimization of pilot bus voltage setpoints and reactive power contributions is performed discretely every 3 hours of simulation time, for a total of eight times within a day, always aiming at loss minimization as per the objective function (3). SVR operation, on the other hand, is carried out continuously as per the parallel scheme of Fig. 2.

Fig. 5 shows the evolution of voltage magnitudes for all buses of Area 1, where it is possible to see a clear distinction between standard operation and the proposed approach.

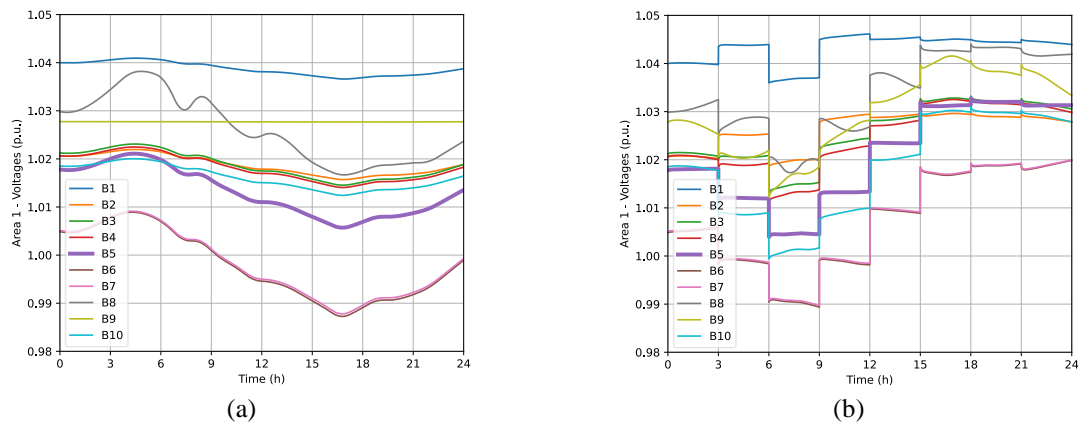

*Figure 5 – Area 1 voltage magnitudes: (a) with no hierarchical control; (b) with the SVR-TVR approach*

In Fig. 5(a), although voltage levels remain within an acceptable range throughout the simulation, they are revealed to be suboptimal in terms of loss reduction. This is proven through Fig. 5(b), where the 3-hourly OPF updates accurately respond to the daily load variation by adjusting all bus voltage levels. During the intervals of no TVR action, the SVR layer ensures that the correct setpoints are maintained, thereby facilitating the goal of loss minimization.

Similarly, Fig. 6 shows the evolution of voltage magnitudes for all buses in Area 2. In Fig. 6(a), the daily behavior is revealed to be predominantly dictated by the wind power patterns in this area, instead of load demand. The influence of such resource is mitigated by the OPF updates as proven through Fig. 6(b). This area is otherwise fairly stable in terms of voltage levels even with hierarchical control, meaning that the potential for network loss reduction comes mostly from optimal management of Area 1 resources.

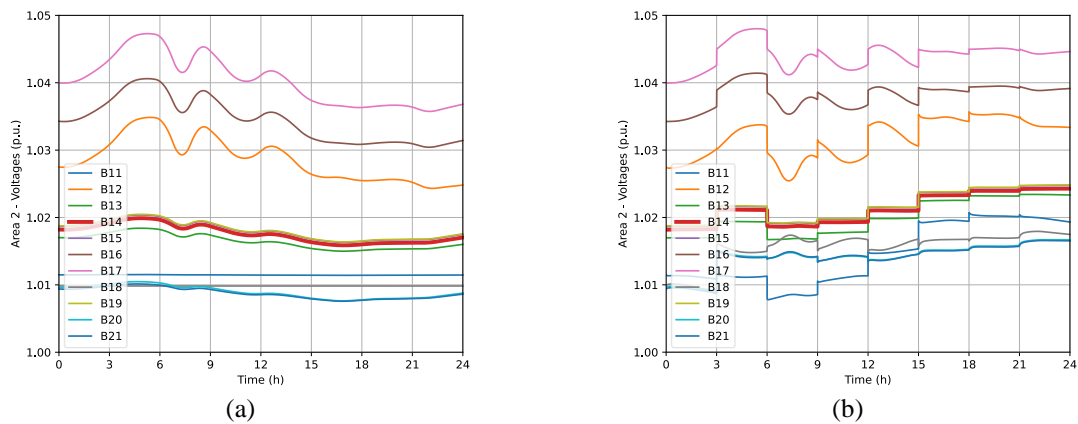

*Figure 6 – Area 2 voltage magnitudes: (a) with no hierarchical control; (b) with the SVR-TVR approach*

Fig. 7 focuses on the daily reactive power contributions from the four hydrogenerators, which are all included in the SVR scheme (G1 and G2 in Area 1, G3 and G4 in Area 2). Fig. 7(a) shows that G1 is typically responsible for a higher share of MVAr contribution than the other three combined, although never surpassing 30 MVAr. On the other hand, Fig. 7(b) reveals that the OPF requires an even higher participation of G1, which now surpasses 40 MVAr during the peak period, as the other machines (and especially G2) are overall spared throughout the simulation. This reinforces the idea that proper management of Area 1 resources is key towards loss minimization in this network.

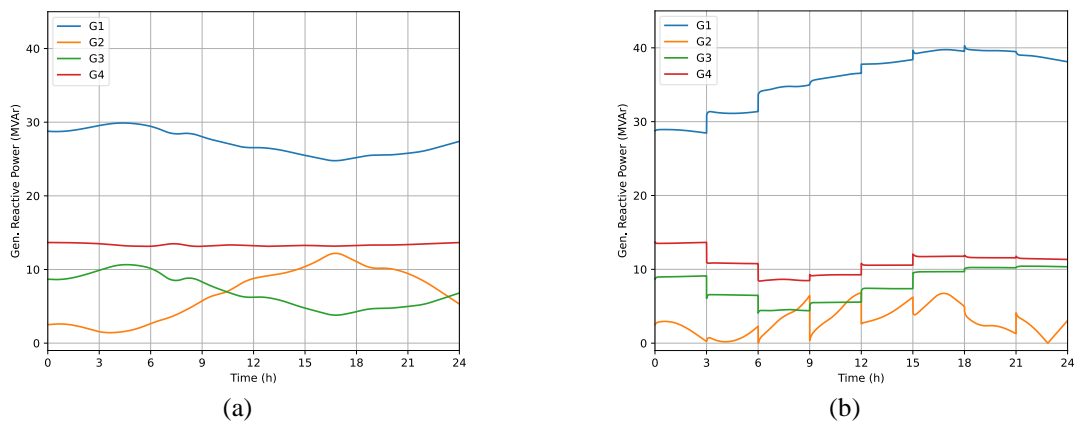

*Figure 7 – Generator MVAr contributions: (a) with no hierarchical control; (b) with the SVR-TVR approach*

Finally, Fig. 8 compares the active power losses obtained from standard network operation and from applying the proposed SVR-TVR approach. The difference in behavior starts as soon as the first OPF update and it only builds up a larger gap from that point on, getting progressively wider as it approaches the peak afternoon period. It is evident from this result that the SVR-TVR adjustments carried out over voltage levels and MVAr contributions had a significant positive impact towards the fulfilment of the OPF objective.

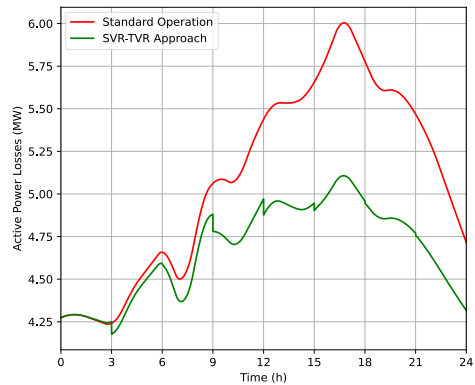

*Figure 8 – Comparison of network active power losses with and without the proposed SVR-TVR approach*

In this scenario, the peak loss reduction corresponds to 800 kW or 13.3% with respect to the standard operation. On average, considering the entire daily profile, loss reduction is in the region of 410 kW or 6.8% for this system. Assuming a day-ahead price of 10 €/MWh, which is reasonably common for the corresponding bidding zone, average cost savings related to network losses reach 4.10 €/h, meaning 100 €/day or 36000 €/year for the TSO. Naturally, such metrics are estimates that can fluctuate depending on demand/generation profiles, device setpoints, contingency events, etc. For the specific subsystem shown in Fig. 3, further tests have consistently provided an average loss reduction of around 6%, also attesting to the merits of the proposed hierarchical control framework.

# 6   Concluding Remarks

This paper presented a combined SVR-TVR framework aimed at improving network performance through minimization of total active power losses given certain system limitations, while simultaneously maintaining adequate bus voltage profiles and reactive power flows within the two control areas through optimal management of VAr resources. The particularities of the studied Norwegian grid subsystem were taken into account by considering a regular day of operation with 24-hour real profiles for load demand and wind power generation patterns.

Results emphasized not only the importance of system-wide hierarchical control for proper utilization of existing grid resources, but also the merits of the SVR-TVR approach proposed for real-time loss minimization as a costs-saving mechanism for TSOs in day-to-day control center operations. An average loss reduction of 6% was obtained from several tests performed in this network, a result which can be replicated in other subsystems under similar conditions.

Future work is intended to expand the scope of dynamic simulation analyses for this system so as to include more complex operational scenarios, such as N-1 contingency events, topology reconfiguration and voltage control through the wind power resources.

## Acknowledgments

The authors gratefully acknowledge the economic support from The Research Council of Norway (RCN) and industry partners through RCN project "ref: 326673 - System Optimization between power producer and grid owners for more efficient system services" (SysOpt).